\documentclass[twocolumn,prl,aps,showpacs]{revtex4}
\usepackage{graphicx}
\usepackage{amsmath}
\usepackage{bm}
\usepackage{epsf,epsfig,graphics}
\usepackage{float}
\usepackage{makeidx}
\usepackage{latexsym}
\usepackage{amssymb}
\usepackage{mathrsfs}
\usepackage{amsbsy}
\usepackage{subfigure}
\usepackage{float}

\usepackage{color}
\usepackage{morefloats}
\DeclareMathAlphabet\mathbfcal{OMS}{cmsy}{b}{n}

\begin{document}

\title{Non-equilibrium thermodynamics and Phase transition of  Ehrenfest urns with interactions}

\author{Chi-Ho Cheng$^{1,}$}
\email{phcch@cc.ncue.edu.tw}
\author{Pik-Yin Lai$^{2,}$}
\email{pylai@phy.ncu.edu.tw}

\affiliation{$^1$Department of Physics, National Changhua University
of Education, Taiwan \\
$^2$Department of Physics and Center for Complex Systems, National Central University, Taiwan}

\date{\today}

\begin{abstract}
Ehrenfest urns with interaction that are connected in a ring is considered as a paradigm model for non-equilibrium  thermodynamics and is shown to exhibit two distinct non-equilibrium steady states (NESS) of uniform  and non-uniform particle distributions.
As the inter-particle attraction varies, a first order non-equilibrium phase transition occurs between these two NESSs characterized by a coexistence regime. The phase boundaries, the NESS particle distributions near saddle points and the associated particle fluxes, average urn population fractions, and the relaxational dynamics to the NESSs are obtained analytically and verified numerically. A generalized non-equilibrium thermodynamics law is also obtained, which explicitly identifies the heat, work, energy and entropy of the system.
\end{abstract}


\maketitle


\section{I. Introduction}
The classic Ehrenfest model \cite{ehrenfest} was introduced to
solve the reversal and Poincare recurrence paradox \cite{huang,poincare}.
It describes a total of $N$ particles in two urns that can randomly jump from one urn
to the other with equal probability.
The system has a Poincare cycle time of $2^N$ \cite{kac}, providing a fundamental link  between reversible microscopic dynamics and irreversible thermodynamics.

Later on, a directional jumping rate between urns was introduced \cite{siegert,klein},
and extensions to multi-urns \cite{kao03,nagler05,clark} were made.
Although there are various modifications
\cite{nagler99,bouchaud,boon,lutsko,schwammle,shiino,frank,chavanis,lipowski02a,lipowski02b,coppex,shim}
of the classic Ehrenfest model,
or even extensions by incorporating non-linear contribution \cite{casas,curado,nobre,frank2005},
particles do not interact or
the  interaction is merely phenomenological.
In fact, pairwise particle interaction
in the same urn has been considered recently in the two-urn model \cite{cheng17}. The interacting two-urn Ehrenfest model can exhibit phase transitions by varying the interaction strength and directional
jumping rate from which the relaxation time and  Poincar\'e cycle can  be derived.
The multi-urn  system with interaction and unbiased  directional
jumping will evolve to  equilibrium and has been shown to exhibit different levels of non-uniformity emerge with the coexistence of uniform and non-uniform
phases \cite{cheng20}.
Contrary to the better understood equilibrium cases, non-equilibrium statistical physics remains challenging, partly due to the lack of well-characterised states. Even for non-equilibrium steady states (NESS), it is difficult to describe non-equilibrium phase transitions between different NESS and their relationship to some microscopic models. For example, selection rules, such as maximal or minimal entropy production principles \cite{nicolis,MaxSbook,Grandy} have been proposed to determine the non-equilibrium states. Yet, universal guiding principles are still lacking.

In this manuscript, we consider  urns with intra-urn interactions connected in a one-dimensional ring with directional jumping rates.
We will show that the system has non-equilibrium steady states in uniform and non-uniform phases
that can coexist. For high directional jumping rates
with appropriate interaction strengths, the steady states become unstable. The phase diagram will be obtained analytically,
and the relaxation dynamics to the NESS will be studied. The relationship between
non-equilibrium thermodynamical variables, such as the internal entropy production rate, the rate of
work done applied to the system, will be shown to obey a generalized thermodynamic law.
Finally, we will demonstrate that, in the coexistence region, the internal
entropy production rate fails to select the favorable steady state.


\section{II. Ehrenfest urns in a ring}
\begin{figure}[tbh]
  \begin{center}
    \includegraphics[width=3in]{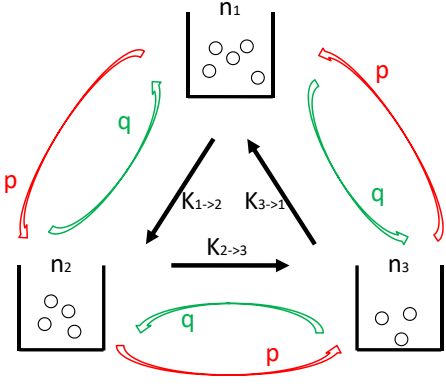}
  \end{center}
  \caption{Schematic diagram of the model. Three urns with particle numbers
    $n_1$, $n_2$, and $n_3$ are connected in a ring. The direct jumping rate in
    anticlockwise (clockwise) direction is $p$ ($q$). $K_{i\rightarrow j}$ represents
    net particle flow rate from the $i$-th to the $j$-th urn.  }
  \label{schematic.eps}
\end{figure}

We consider three urns as illustrated in Fig.\ref{schematic.eps}
as this already captures the non-equilibrium physics.
The state of the system is labeled by $\vec n = (n_1,n_2,n_3)$ where $n_i$ is the number of particles in the $i$-th urn with $n_1 + n_2 + n_3 = N$, the fixed total particle number.
Similar to previous models \cite{cheng17,cheng20}, we include a pairwise  attractive (repulsive) interaction  with negative (positive) energy $J$  for particles in the same urn. Particles in different urns do not interact.
A particle in the $i$-th urn (initial state $\vec n$) jumps to
the $j$-th urn (final state $\vec m$) with corresponding transition probability
\begin{eqnarray}  \label{tnm}
  T_{\vec m,\vec n} &=&
  \frac{1}{{\rm e}^{-\frac{g}{N}(n_i-n_j-1)}+1}
\end{eqnarray}
where $m_i = n_i-1$ and $m_j = n_j+1$.
$g\equiv N J \beta$ where $\beta$ is the inverse of effective temperature.
Without interaction ($g=0$), we have $T_{\vec m, \vec n} = \frac{1}{2}$.
Next, a jumping rate is introduced such that the probability
of anticlockwise (clockwise) direction is $p$ ($q$).
For the sake of convenience, $p+q=1$ is imposed for
which only changes the time scale.

After $s$ steps from the initial state, the probability
of the state $\vec n$ is denoted by $\rho(\vec n,s)$.
The master equation
from the $s$-th to the $(s+1)$-th time step can be written as
\begin{eqnarray} \label{recurrence01}
  \rho(\vec n,s+1) - \rho(\vec n, s)
  = \sum_{\vec m} ( W_{\vec n,\vec m} \rho(\vec m,s) - W_{\vec m, \vec n} \rho(\vec n,s) )
\end{eqnarray}
where
\begin{eqnarray} \label{w01}
W_{\vec m, \vec n} = \frac{n_i}{N} p T_{\vec m, \vec n}
\end{eqnarray}
holds if the particle jumps from the $i$-th to the $j$-th urn is anticlockwise, {\rm i.e.},
$(i,j)=(1,2),(2,3),(3,1)$, and
\begin{eqnarray} \label{w02}
  W_{\vec m, \vec n} = \frac{n_i}{N} q T_{\vec m, \vec n}
\end{eqnarray}
if $(i,j)=(2,1),(1,3),(3,2)$, which represents clockwise jumps.

Finally, the net particle flow rate from the $i$-th to the $j$-th urn
(see Fig.\ref{schematic.eps} for illustration) is given by
\begin{eqnarray} \label{netflow01}
  K_{i\rightarrow j}(s) \equiv N \sum_{\vec n} ( W_{\vec m,\vec n} - W_{\vec m',\vec n} ) \rho(\vec n,s)
\end{eqnarray}
where $m_i=n_i-1$, $m_j=n_j+1$, and $m'_i=n_i+1$, $m'_j=n_j-1$.

\section{III. Equilibrium states}
If $\lim_{s\rightarrow\infty} \rho(\vec n,s)$ exists,
it defines the steady state $\rho^{\rm ss}(\vec n)$.
Taking the limit $s\rightarrow\infty$ in Eq.(\ref{recurrence01}) would lead to an
equation in which $\rho^{\rm ss}(\vec n)$ satisfies. No closed form exists in general.
But for the  case of $p=q=\frac{1}{2}$, an analytic expression for $\rho^{\rm ss}$ can be obtained
\begin{eqnarray} \label{rhoeq}
  \rho^{\rm ss}(\vec n) = \frac{N!}{n_1! n_2! n_3!}
      {\rm e}^{-\frac{g}{2 N}(n_1^2 + n_2^2 + n_3^2)}
\end{eqnarray}
up to a normalization constant. This steady state is also the equilibrium state,
because it satisfies the condition of detailed balance
\begin{eqnarray} \label{db07}
  W_{\vec m, \vec n} \rho^{\rm ss}(\vec n)
  = W_{\vec n, \vec m} \rho^{\rm ss}(\vec m)
\end{eqnarray}
which can be verified by direct substitution. Results for the general case of $M$ urns at equilibrium can be found in Ref.\cite{cheng20}. For the  three urns case here with
 the constraint $n_1+n_2+n_3=N$, one can define the population fraction with $x_i\equiv n_i/N$ and rewrite $\rho^{\rm ss}$ in terms of ${\vec x}\equiv (x_1,x_2)$ (two independent variables),
$\rho^{\rm ss}(\vec x)={\rm exp}\{N f({\vec x})
- \frac{1}{2}\log[(2\pi N)^2 x_1 x_2 (1-x_1-x_2)] + O(N^{-1}) \}$ in large $N$ limit,
where
\begin{eqnarray}
  f({\vec x}) &=& -x_1\ln x_1-x_2\ln x_2  \nonumber \\
  && -(1-x_1-x_2)\ln(1-x_1-x_2) \nonumber \\
  && -\frac{g}{2}(x_1^2+x_2^2+(1-x_1-x_2)^2)
\end{eqnarray}
The saddle point approximation \cite{saddle} gives asymptotic form
\begin{eqnarray}
\rho^{\rm ss}(\vec x) \propto
{\rm e}^{ N f(\vec x^{\rm sp})
+ \frac{N}{2} \sum_{i,j} \partial_{i j} f({\vec x}^{\rm sp}) (x_i-x_i^{\rm sp})(x_j-x_j^{\rm sp}) }
\end{eqnarray}
where ${\vec x}^{\rm sp}$ is the saddle point(s) satisfying
$\partial_1 f = \partial_2 f = 0$,
$\partial_{11} f < 0$, and
$\partial_{11} f \partial_{22} f-(\partial_{12}f)^2 > 0$. This condition leads to
\begin{eqnarray} \label{mfeq01}
  x_1^{\rm sp} {\rm e}^{g x_1^{\rm sp}}
  = x_2^{\rm sp} {\rm e}^{g x_2^{\rm sp}}
  = x_3^{\rm sp} {\rm e}^{g x_3^{\rm sp}}
\end{eqnarray}
The solutions at different coupling constant $g$ are shown in the data of $p=0.5$ in Fig.2(a).
The steady net particle flow at equilibrium from Eq.(\ref{netflow01}) reads
\begin{eqnarray}
  \frac{K^{\rm ss}_{i\rightarrow j}}{N}
  &=& \frac{ x_i^{\rm sp} {\rm e}^{g x_i^{\rm sp}} - x_j^{\rm sp} {\rm e}^{g x_j^{\rm sp}}}
  {2({\rm e}^{g x_i^{\rm sp}} + {\rm e}^{g x_j^{\rm sp}})}
   + O(1/N)
\end{eqnarray}
which gives $K^{\rm ss}_{1\rightarrow 2}
= K^{\rm ss}_{2\rightarrow 3}= K^{\rm ss}_{3\rightarrow 1} = 0$
from Eq.(\ref{mfeq01}), in large $N$ limit.
At equilibrium, there is no net particle flow between any two urns.
In addition, it is also easy to see from Eqs.(\ref{netflow01}) and (\ref{db07}) that for $p=q=\frac{1}{2}$, all fluxes $K^{\rm ss}_{i\rightarrow j}=0$ at equilibrium.
On the other hand,  or the non-equilibrium case of $p\neq q$, there can be non-vanishing circulating fluxes  as in other general NESS systems \cite{aopingPNAS,Chiang2017,Chang2021}.

\section{IV. Uniform and non-uniform non-equilibrium steady states}
So far, the recurrence relation in Eq.(\ref{recurrence01}) cannot be solved analytically
even for NESS.
In this section, we will transform Eq.(\ref{recurrence01}) into the Fokker-Planck equation.
Let the (physical) time $t= \frac{\tau_1}{N} s $, where
$\tau_1$ is the time scale of each single step from $s$ to $s+1$.
Replace $\rho(\vec n,s+1)-\rho(\vec n,s)$ by
$\rho(\vec n,t+\frac{\tau_1}{N})-\rho(\vec n,t)
= \frac{\tau_1}{N}\frac{\partial\rho}{\partial t} + O((\frac{\tau_1}{N})^2)$.
Eq.(\ref{recurrence01}) can be rewritten as
\begin{eqnarray} \label{recurrence02}
  \frac{\tau_1}{N}\frac{\partial\rho(\vec n,t)}{\partial t}
  = \sum_{\vec m} ( W_{\vec n,\vec m} \rho(\vec m,t) - W_{\vec m, \vec n} \rho(\vec n,t) )
\end{eqnarray}
Substituting Eqs.(\ref{tnm}), (\ref{w01})-(\ref{w02}) into Eq.(\ref{recurrence02})
gives
\begin{eqnarray} \label{km}
  && \frac{\tau_1}{N}\frac{\partial \rho(\vec x,t)}{\partial t}  \nonumber \\
  &=& p \sum_{k=1}^\infty \frac{1}{k! N^k}
  \left(\frac{\partial}{\partial x_1}-\frac{\partial}{\partial x_2} \right)^k
  [\frac{x_1}{{\rm e}^{-g(x_1-x_2)}+1} \rho(\vec x,t)]  \nonumber \\
  && + q \sum_{k=1}^\infty \frac{1}{k! N^k}
  \left(-\frac{\partial}{\partial x_1}+\frac{\partial}{\partial x_2} \right)^k
       [\frac{x_2}{{\rm e}^{g(x_1-x_2)}+1} \rho(\vec x,t)] \nonumber \\
       && + [\rm cyclic\ terms]
\end{eqnarray}
which is known as the Kramers-Moyal expansion \cite{kramers,moyal}. From now on, we
take $\tau_1 = 1$ for convenience.

\begin{figure}[tbh]
  \begin{center}
    \includegraphics[width=3in]{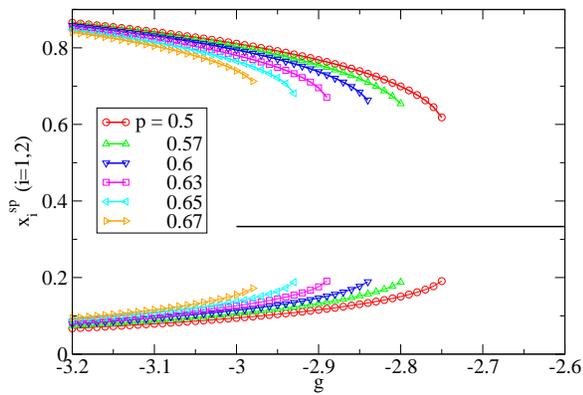}
  \end{center}
  \vspace{-5pt}
  \caption{Occupation fraction at the stable saddle point in the $i$-th urn,
  $x_i^{\rm sp}$, as a function of coupling constant $g$ for different $p$.
  Up to the cyclic permutation, we consider $x_1^{\rm sp}$ the largest fraction, and
  $x_2^{\rm sp}$ is the occupation fraction in the next urn along the $p$ direction.
  The remaining $x_3^{\rm sp}=1-x_1^{\rm sp}-x_2^{\rm sp}$.
        Solid line (black) represents the stable saddle point
    $x_1^{\rm sp}=x_2^{\rm sp}=x_3^{\rm sp}=\frac{1}{3}$
    (uniform distribution) shared by all values of $p$. }
  \label{xvsg.eps}
\vspace{25pt}
\end{figure}

If we further keep terms up to $O(1/N^2)$,
Eq.(\ref{km}) becomes the Fokker-Planck equation
\begin{eqnarray} \label{fkeq02}
   \frac{\partial \rho(\vec x,t)}{\partial t}
  &=& -\sum_{i=1}^2\frac{\partial}{\partial x_i}[A_i(\vec x) \rho(\vec x;t)] \nonumber \\
   && + \frac{1}{2 N}\sum_{i,j=1}^2 \frac{\partial^2}{\partial x_i \partial x_j}
  [B_{ij}(\vec x) \rho(\vec x;t)]
\end{eqnarray}
where
\begin{widetext}
\begin{eqnarray}
  A_1(\vec x) &=& -\frac{p x_1}{{\rm e}^{-g(x_1-x_2)}+1}
  + \frac{q x_2}{{\rm e}^{g(x_1-x_2)}+1}
  - \frac{q x_1}{{\rm e}^{-g(2 x_1+x_2-1)}+1}
  + \frac{p (1-x_1-x_2)}{{\rm e}^{g(2 x_1+x_2-1)}+1}  \\
  A_2(\vec x) &=& -\frac{q x_2}{{\rm e}^{-g(x_2-x_1)}+1}
  + \frac{p x_1}{{\rm e}^{g(x_2-x_1)}+1}
  - \frac{p x_2}{{\rm e}^{-g(2 x_2+x_1-1)}+1}
  + \frac{q (1-x_1-x_2)}{{\rm e}^{g (2 x_2+x_1-1)}+1}  \\
  B_{11}(\vec x) &=& \frac{p x_1}{{\rm e}^{-g(x_1-x_2)+1}}
  + \frac{q x_2}{{\rm e}^{g(x_1-x_2)}+1}
  + \frac{q x_1}{{\rm e}^{-g(2 x_1+x_2-1)}+1}
  + \frac{p(1-x_1-x_2)}{{\rm e}^{g(2 x_1+x_2-1)}+1}   \\
  B_{22}(\vec x) &=& \frac{q x_2}{{\rm e}^{-g(x_2-x_1)}+1}
  + \frac{p x_1}{{\rm e}^{g(x_2-x_1)}+1}
  + \frac{p x_2}{{\rm e}^{-g(2 x_2+x_1-1)}+1}
  + \frac{q(1-x_1-x_2)}{{\rm e}^{g(2 x_2+x_1-1)}+1} \\
  B_{12}(\vec x) &=& B_{21}(\vec x) =
  -\frac{p x_1}{{\rm e}^{-g(x_1-x_2)}+1} -\frac{q x_2}{{\rm e}^{g(x_1-x_2)}+1}
\end{eqnarray}
\end{widetext}
The WKB approximation \cite{wkb,kubo,dykman,elgartv,assaf} yields the saddle points \cite{saddle}
${\vec x}^{\rm sp}=(x_1^{\rm sp},x_2^{\rm sp})$ as
\begin{equation}
  A_1({\vec x}^{\rm sp}) = 0, \quad   A_2({\vec x}^{\rm sp}) = 0\label{A1A2}
\end{equation}
whose solutions at different $g$ and $p$ are shown in Fig.\ref{xvsg.eps}.
The physical meaning of Eq.(\ref{A1A2}) is that
$K^{\rm ss}_{1\rightarrow 2}=K^{\rm ss}_{2\rightarrow 3}=K^{\rm ss}_{3\rightarrow 1}=K^{\rm ss}$,
i.e., a constant non-zero cyclic flux of net particle along the ring which can be calculated
from Eq.(\ref{netflow01}) as
\begin{eqnarray}
  \frac{K^{\rm ss}}{N}
  &=& \frac{ p x_1^{\rm sp} {\rm e}^{g x_1^{\rm sp}} -q x_2^{\rm sp} {\rm e}^{g x_2^{\rm sp}}}
  {{\rm e}^{g x_1^{\rm sp}} + {\rm e}^{g x_2^{\rm sp}}}
   \label{Kss}
\end{eqnarray}

\begin{figure}[tbh]
  \begin{center}
    \includegraphics[width=3in]{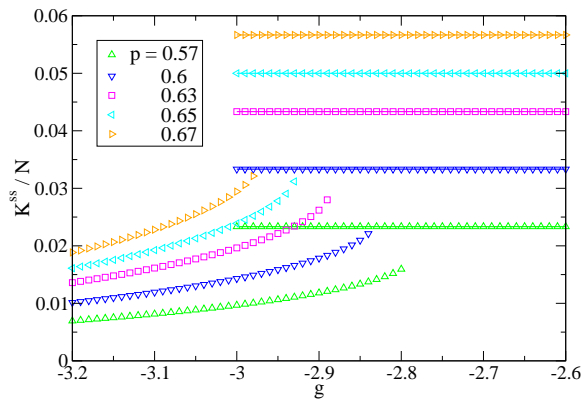}
  \end{center}
  \vspace{-5pt}
  \caption{Net particle flow of the uniform and non-uniform NESS
    as a function of coupling constant $g$ for different $p$.
    Symbols with and without lines represent uniform and non-uniform
    distributions, respectively. }
  \label{kssvsg.eps}
\vspace{25pt}
\end{figure}

For uniform NESS ($x_1=x_2=\frac{1}{3}$), one obtains
$K^{\rm ss}_{\rm u}= {N\over 6}(p-q)$
and, for non-uniform NESS, $K^{\rm ss}_{\rm nu}$ can be computed using the non-uniform saddle point from Eqs.(\ref{A1A2}). The $K^{\rm ss}_{\rm u}$ and  $ K^{\rm ss}_{\rm nu}$ NESS fluxes as a function of $g$ at different $p$ are shown in Fig.\ref{kssvsg.eps} indicating that the particle flux of the uniform NESS is always significantly larger than that of the non-uniform NESS. Notice that there is a coexistence region of uniform and non-uniform saddle points.

\vspace{25pt}

\vspace{25pt}

The NESS can be further analysed by expanding around the saddle point, i.e. using
$A_i(\vec x) \simeq \sum_j \partial_j A_i(\vec x^{\rm sp})(x_j-x_j^{\rm sp})
\equiv \sum_j a_{ij}(x_j-x_j^{\rm sp})$ and
$B_{ij}(\vec x) \simeq B_{ij}(\vec x^{\rm sp})=b_{ij}$, the steady state particle distribution is
\begin{eqnarray}
  \rho^{\rm ss}(\vec x) \propto \exp[ N \sum_{i,j=1}^2 c_{ij} (x_i-x_i^{\rm sp}) (x_j-x_j^{\rm sp})]
\end{eqnarray}
where the matrix $\bf c$ is uniquely determined by the Lyapunov equations
${\bf a} {\bf c}^{-1} + {\bf c}^{-1} {\bf a}^{\rm t} = 2 {\bf b}$ (See Appendix A for details).
The detailed balance condition can be transformed into
${\bf c} = {\bf b}^{-1} {\bf a}$ (See Appendix B for details).

\begin{figure}[tbh]
  \begin{center}
  \includegraphics[width=3in]{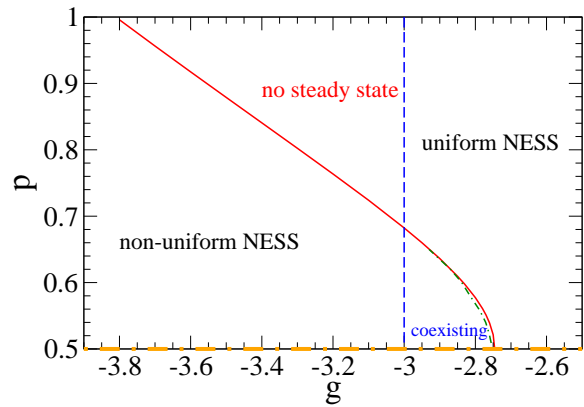}
  \end{center}
  \caption{Phase diagram of the interacting Ehrenfest model of three urns
    connected in a ring. There are four regions which represent
    uniform NESS (particles uniformly distributed in three urns),
    non-uniform  NESS,    coexistence (both uniform and non-uniform NESSs are stable),
    and no steady state. The stability boundaries of the uniform NESS and non-uniform NESS are denoted by the vertical dashed line and solid curve respectively. The first order phase transition boundary in the coexistence regime is denoted by the dashed-dotted curve (see Fig.\ref{phasezoom.eps} in Appendix C for a magnification).}
  \label{phase1.eps}
  \end{figure}

Obviously $x_1^{\rm sp}=x_2^{\rm sp}=x_3^{\rm sp}=\frac{1}{3}$
is always a saddle point of uniform population fraction. At this uniform NESS, we have
\begin{eqnarray}
  {\bf a} &=& -\frac{1}{2} \left(\begin{array}{cc}
  1+p+\frac{g}{2} & p-q  \\
  q-p & 1+q+ \frac{g}{2} \\
  \end{array}
  \right)  \label{uni-a} \\
  {\bf b} &=& \frac{1}{6} \left(\begin{array}{cc}
    2 & -1 \\
    -1 & 2 \\
  \end{array}
  \right)  \label{uni-b}
\end{eqnarray}
which gives
\begin{eqnarray}
  {\bf c} = -\frac{g+3}{2} \left(\begin{array}{cc}
    2 & 1 \\
    1 & 2 \\
    \end{array}\right)
\end{eqnarray}
Hence its stability requires $g > -3$.
The stable region for the non-uniform phase can be determined analytically in a similar manner
and the results agree with the analytic ones of the phase boundary in Fig.\ref{phase1.eps}.
In fact, the non-equilibrium physics of the system can be summarized by the phase diagram in Fig.\ref{phase1.eps} whose phase boundaries can be determined analytically
(see Appendix C for derivations).
As the particle interaction
is strongly repulsive (positively large $g$), the particles are uniformly
distributed in every urn. On the other hand, the particles ``prefer'' to stay in the same
urn (non-uniform distribution) if they are strongly attractive (negatively
large $g$). In between, for low jumping rate, $1-p_{\rm s}<p<p_{\rm s}$ $(p_{\rm s}\simeq 0.6823)$,
there is a  coexistence region where both uniform and non-uniform distribution
are locally stable.
There is a first order non-equilibrium phase transition between the uniform and non-uniform NESSs whose transition value of $g$ can be determined from the analytic result of $ K^{\rm ss}_{\rm u} $ and $ K^{\rm ss}_{\rm nu} $ (see Appendix C for details) together with the results of mean steady state flux. The latter has to be determined numerically using (\ref{recurrence01}) .
The first order phase transition line (dashed-dotted curve) is also shown in Fig.\ref{phase1.eps}.
It is close to the stability line of the non-uniform NESS
(for magnification, see Fig.\ref{phasezoom.eps} in Appendix C).
As the jumping rate becomes higher, i.e. $p > p_{\rm s}$ (or $p < 1-p_{\rm s}$), the system is far from equilibrium and steady states do no longer exist.

\section{V. Relaxation to steady states}
\begin{figure}[tbh]
  \begin{center}
   \includegraphics[width=3in]{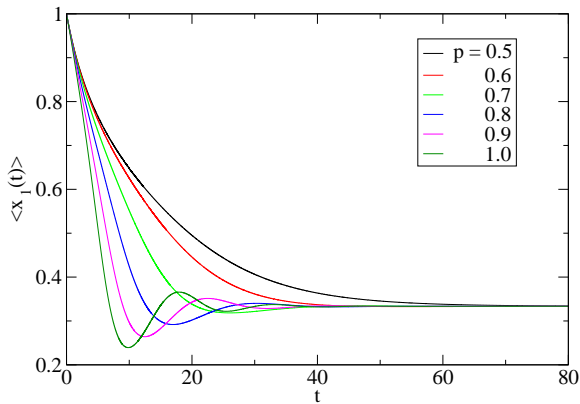}
  \end{center}
  \caption{$\langle x_1(t)\rangle$ as a function of $t$ at different $p$  with $g$=-2.5.
  The initial state is $\vec x(0)=(1,0,0)$
  and the steady state is uniform. The result is     numerically solved from Eq.(\ref{recurrence01}) for $N=300$. }
  \label{gn2d5_n1_vs_t.eps}
\end{figure}

In this section, we studied how the system evolves to its steady states.
When the system is initially away from its steady state, it will relax (thermalized) towards the nearby stable NESS whose dynamics near the NESS is determined by the eigenvalues of ${\bf a}$.
From Eq.(\ref{fkeq02}),
and expand around the saddle point $\vec x^{\rm sp}$, we get the evolution of
the average of particle numbers in the first and second urn as
\begin{eqnarray}
\frac{d}{dt}\left( \begin{array}{c}
                   \langle x_1(t) \rangle \\
                   \langle x_2(t) \rangle \\
            \end{array} \right)
 = {\bf a}
 \left( \begin{array}{c}
        \langle x_1(t) \rangle - x_1^{\rm sp}\\
        \langle x_2(t) \rangle - x_2^{\rm sp} \\
        \end{array} \right)
\end{eqnarray}
At uniform phase, $x_1^{\rm sp}=x_2^{\rm sp}=x_3^{\rm sp}=\frac{1}{3}$,
its solution is
\begin{widetext}
\begin{eqnarray} \label{x1}
\langle x_1(t) \rangle &=& \frac{1}{3}
+  \left\{
(x_1(0)-\frac{1}{3}) \cos(\frac{2\pi t}{\tau_{\rm osc}})
- \frac{1}{\sqrt{3}} (x_2(0)-x_3(0)) \sin(\frac{2\pi t}{\tau_{\rm osc}})
\right\}  {\rm e}^{-t/\tau_{\rm R}}
\end{eqnarray}
\end{widetext}
describing the decaying process with oscillation, with
the relaxation time
\begin{eqnarray}
\tau_{\rm R} = \frac{4}{g+3}
\end{eqnarray}
and the oscillation period
\begin{eqnarray}
\tau_{\rm osc} = \frac{8\pi}{\sqrt{3} |p-q| }
\end{eqnarray}
$x_2(t)$ and $x_3(t)$ can be obtained by making cyclic transformation to Eq.(\ref{x1}).
Near equilibrium ($|p-q| \ll 1$), $\tau_{\rm osc} \gg \tau_{\rm R}$ ,
then the solution is simplified as
\begin{eqnarray}
\langle x_i(t) \rangle = \frac{1}{3} + (x_i(0) - \frac{1}{3}) {\rm e}^{-t/\tau_{\rm R}}
\end{eqnarray}
and the damped oscillation is not prominent.
Fig.\ref{gn2d5_n1_vs_t.eps} shows the relaxation towards the uniform NESS for different $p$,
with $g=-2.5$ at which only the uniform NESS is stable. By increasing $p$ from 0.5 to 1.0, it can be seen from the direct numerical calculation by Eq.(\ref{recurrence01}), starting from the non-uniform initial state $\vec x(0)=(1,0,0)$,
the relaxation time keep almost unchanged and the oscillation in the occupation gradually appears.

\begin{figure}[tbh]
  \begin{center}
   \includegraphics[width=3in]{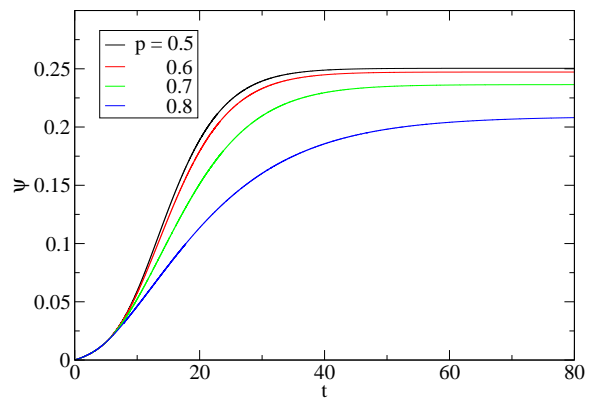}
  \end{center}
  \caption{Non-uniformity parameter $\psi$ as a function of $t$
  for different $p$ when $g$=-3.5.
  The initial state is $\vec x(0)=(\frac{1}{3},\frac{1}{3},\frac{1}{3})$
  and the steady state is non-uniform. The result is
    numerically solved from Eq.(\ref{recurrence01}) for $N=300$.}
  \label{gn3d5_psi_vs_t.eps}
\end{figure}

\begin{figure}[tbh]
  \begin{center}
   \includegraphics[width=3in]{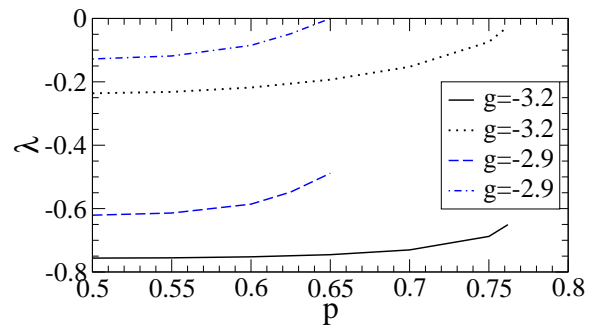}
  \end{center}
  \caption{The eigenvalues of $\bf a$ plotted against $p$, for the non-uniform NESS ($g=-3.2$)
  and in the coexisting NESS ($g=-2.9$). The relaxation to the non-uniform NESS
  is always purely relaxational.}
  \label{lambdanuNESS.eps}
\end{figure}

When $g=-3.5$, the system stays at the non-uniform phase.
To quantify the degree of non-uniformity, we define
 \begin{eqnarray}
\psi \equiv \frac{1}{6} \langle (x_1-x_2)^2 +(x_2-x_3)^2 +(x_3-x_1)^2\rangle
\end{eqnarray}
as the ``non-uniformity" parameter \cite{cheng20}.
It is almost zero for uniform phase
and becomes larger for higher non-uniformity.
The relaxation towards the non-uniform NESS  is illustrated by $\psi(t)$  with $g=-3.5$ in Fig.\ref{gn3d5_psi_vs_t.eps}, showing a pure relaxation behavior.
Starting from the uniform initial state $\vec x(0)=(\frac{1}{3},\frac{1}{3},\frac{1}{3})$, the system for different $p$ all relax to the non-uniform NESS and saturates at a high non-uniformity. This can be understood in terms of the eigenvalues of the matrix $\bf a$ at the non-uniform state which are always real and negative as shown in Fig.\ref{lambdanuNESS.eps}.

\section{VI. Non-equilibrium thermodynamics}
We further examine the relationship between various thermodynamical quantities, first for general non-equilibrium states and then for the NESS cases. The Boltzmann (Gibbs, Shannon) entropy
of the system is given by
\begin{eqnarray}
S = -\sum_{\vec n} \rho(\vec n,t)
\log \left( \rho(\vec n,t)/\frac{N!}{n_1! n_2! n_3!}
     \right)
\end{eqnarray}
where the multiplication factor $\frac{N!}{n_1! n_2! n_3!}$ is due to the degeneracy of $\rho(\vec n,t)$ \cite{huang}. Applying Eq.(\ref{recurrence01}),
the entropy production rate above can be written as
\begin{widetext}
\begin{eqnarray} \label{dSdt01}
\frac{dS}{dt} &=& -\sum_{\vec n, \vec m}
(W_{\vec n, \vec m} \rho(\vec m,t) - W_{\vec m, \vec n} \rho(\vec n,t))
\log\left(\rho(\vec n,t)/\frac{N!}{n_1! n_2! n_3!}\right) \nonumber \\
&=& \frac{N}{2} \sum_{\vec n, \vec m}
(W_{\vec n, \vec m} \rho(\vec m,t) - W_{\vec m, \vec n} \rho(\vec n,t))
\log\left( \frac{W_{\vec n, \vec m}\rho(\vec m,t)}{W_{\vec m, \vec n}\rho(\vec n,t)} \right)  \nonumber \\
&& +\frac{N}{2} \sum_{\vec n, \vec m}
(W_{\vec n, \vec m} \rho(\vec m,t) - W_{\vec m, \vec n} \rho(\vec n,t))
\log\left( \frac{W_{\vec m,\vec n}}{W_{\vec n,\vec m}}
\frac{\frac{N!}{n_1! n_2! n_3!}}{\frac{N!}{m_1! m_2! m_3!}} \right) \nonumber \\
&=& \frac{d_{\rm i} S}{dt} + \frac{d_{\rm e} S}{dt}
\end{eqnarray}
\end{widetext}
where the first term is the internal entropy production rate \cite{schnakenberg}, which  is positive-definite and only vanishes when the system is at equilibrium (Eq.(\ref{db07})).
It is the entropy produced during the irreversible process \cite{prigogine}. The second term refers to
the entropy production rate for the reversible process \cite{lebon} into the system.
In the following, we will show that
\begin{eqnarray} \label{dSdt02}
\frac{d_{\rm e} S}{dt} = \beta \frac{dE}{dt} + \beta \frac{dW}{dt}
\end{eqnarray}
where $\frac{dE}{dt}$ and $\frac{dW}{dt}$ are the rate of change of system energy
and the rate of work done by the system, respectively. From the first law of thermodynamics
(conservation law of energy), $T d_{\rm e} S$ can be identified as $dQ$, the heat flow to the system
from the environment. In general, during thermalization process, $dS \geq \beta dQ = \beta dE + \beta dW$.

Using Eqs.(\ref{w01})-(\ref{w02}), when the particle jumps from the $i$-th to the $j$-th urn,
corresponding to the transition from state $\vec n$ to $\vec m$,
\begin{eqnarray}
\frac{W_{\vec m, \vec n}}{W_{\vec n, \vec m}} = \frac{p}{q} \frac{n_i}{n_j+1}
{\rm e}^{\frac{g}{N}(n_i-n_j-1)}
\end{eqnarray}
if the jump is in anti-clockwise direction. Otherwise, in clockwise direction,
\begin{eqnarray}
\frac{W_{\vec m, \vec n}}{W_{\vec n, \vec m}} = \frac{q}{p} \frac{n_i}{n_j+1}
{\rm e}^{\frac{g}{N}(n_i-n_j-1)}
\end{eqnarray}
Then
\begin{widetext}
\begin{eqnarray}
\frac{d_{\rm e} S}{dt}
&=& \frac{N}{2}
\sum_{\vec n, \vec m} (W_{\vec n, \vec m} \rho(\vec m,t) - W_{\vec m, \vec n} \rho(\vec n,t) )
\frac{g}{N}(n_i-n_j-1)
\nonumber \\
&& + \frac{N}{2}
\sum_{\vec n}{\sum_{\vec m}}^{\rm ac} (W_{\vec n, \vec m} \rho(\vec m,t) - W_{\vec m, \vec n} \rho(\vec n,t))
\log(\frac{p}{q})
+ \frac{N}{2}
\sum_{\vec n}{\sum_{\vec m}}^{\rm c} (W_{\vec n, \vec m} \rho(\vec m,t) - W_{\vec m, \vec n} \rho(\vec n,t))
\log(\frac{q}{p})
\nonumber \\
&=& \sum_{\vec n, \vec m} g [ n_j-(n_i-1)]
W_{\vec m, \vec n} \rho(\vec n,t)
 - N \log(\frac{p}{q}) \sum_{\vec n} {\sum_{\vec m}}^{\rm ac}
(W_{\vec m, \vec n}\rho(\vec n,t)-W_{\vec n, \vec m}\rho(\vec m,t))
\end{eqnarray}
\end{widetext}
where ac (c) stands for anti-clockwise (clockwise) direction.
The first term is the rate of change of energy $\beta \frac{dE}{dt}$
and the second term is equal to the rate of work done which can be written as
\begin{eqnarray} \label{dWdt}
\beta \frac{dW}{dt} &=& -\beta \mu (K_{1\rightarrow 2} + K_{2\rightarrow 3} + K_{3\rightarrow 1})
\end{eqnarray}
where $\mu\equiv \beta^{-1} \log (\frac{p}{q})$ is the effective chemical potential difference to actively drive the particle
from one urn to another, and the natural boundary condition is assumed.
Here Eq.(\ref{dSdt02}) is proved, and hence a more general thermodynamic law
\begin{equation}
dS=d_{\rm i}S+\beta dE+ \beta dW  \label{Law}
\end{equation}
is derived. Note that Eqs. (\ref{dWdt}) and (\ref{Law}) hold
even for general non-equilibrium (non-steady state) processes.

\begin{figure}[tbh]
  \begin{center}
  \includegraphics[width=3in]{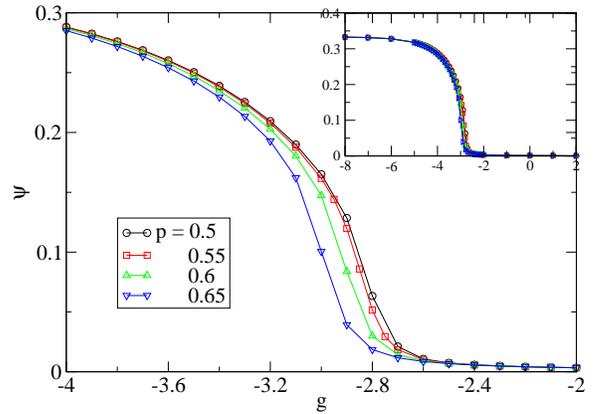}
  \end{center}
  \caption{Non-uniformity parameterat steady states $\psi^{ss}$
    as a function of coupling constant $g$ for different $p$.
    Inset: The same plot with wider range of $g$.
    The result is obtained numerically by Eq.(\ref{diSdt}) with $N=300$. }
  \label{g_vs_psi.eps}
\end{figure}

\begin{figure}[tbh]
  \begin{center}
  \includegraphics[width=3in]{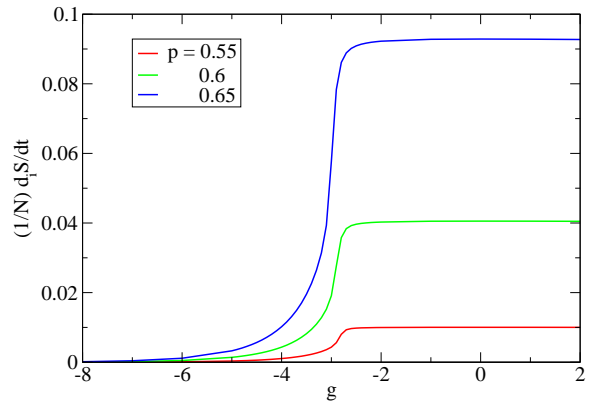}
  \end{center}
  \caption{Internal entropy production rate
     $\frac{1}{N} \frac{d_{\rm i} S}{dt}$ at steady states
    as a function of coupling constant $g$ for different $p$. The result is
    obtained numerically by Eq.(\ref{diSdt}) with $N=300$. }
  \label{g_vs_dW.eps}
\end{figure}

\begin{figure}[tbh]
  \begin{center}
    \includegraphics[width=3in]{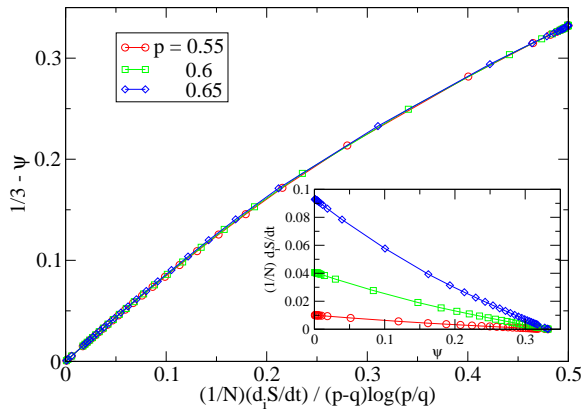}
  \end{center}
  \caption{The relationship between $(\frac{1}{3}-\psi)$
    and $\frac{1}{N}\frac{d_{\rm i}S}{dt}\frac{1}{(p-q)\log(p/q)}$
    for different $p$ at steady states. All data are collapsed into a single curve.
    Inset: The internal entropy production rate $\frac{1}{N}\frac{d_{\rm i}S}{dt}$
    as a function of non-uniformity parameter $\psi$ for different $p$ at steady states.
    The result is
    obtained numerically with $N=300$.    }
  \label{diS2_vs_psi2.eps}
\end{figure}

For $p=q$, the system is at equilibrium and
$\frac{dS}{dt}=\frac{d_{\rm i}S}{dt}=\frac{dQ}{dt}=\frac{dE}{dt}=\frac{dW}{dt}=0$.
That is, all macroscopic thermodynamic quantities do not change with time.
For  $p\neq q$ under NESS, since the system energy and entropy are functionals of the probability distribution and hence  are time independent, thus one has $\frac{dS}{dt}=\frac{dE}{dt}=0$.
Using Eqs.(\ref{dWdt})-(\ref{Law}),
\begin{eqnarray} \label{diSdt}
\frac{d_{\rm i} S}{dt} &=& -\beta\frac{dW}{dt}= -\beta\frac{dQ}{dt}= 3 K^{\rm ss} \log(\frac{p}{q})
\end{eqnarray}
which is a positive constant, corresponding to the house-keeping heat production rate to maintain the NESS.
All the work done ($-dW$) to the system is dissipated (measured by the internal EP
$d_{\rm i}S$)
into heat energy ($-dQ$). Furthermore, the more non-uniform is the NESS (Fig.\ref{g_vs_psi.eps}),
the less is the particle flow (Fig.\ref{kssvsg.eps}),
and hence the less internal EP (Fig.\ref{g_vs_dW.eps}).
Since the internal EP for the non-uniform phase is always lower,
the maximal EP principle could not be used to select the favorable state in the co-existence region.
The first order non-equilibrium phase transition  between the uniform and non-uniform NESS can also be observed by examining the internal EP rate as the particle attraction varies. Fig.\ref{g_vs_dW.eps} shows a sharp drop near some threshold as  $g$ decreases and signifying a first order transition from the high internal EP uniform NESS to the low internal EP non-uniform NESS.
Interestingly, there is a connection between
the internal EP rate and the non-uniformity in the NESS.
As shown in Fig.\ref{diS2_vs_psi2.eps}, when the relationship between $\frac{1}{3} - \psi^{\rm ss}$
and $\frac{1}{N}\frac{d_{\rm i}S}{dt}/(p-q)\log(\frac{p}{q})$ are plotted,
all data with different $p$ are collapsed into a single curve. It implies the relation
\begin{eqnarray}
\left.\frac{d_{\rm i}S}{dt}\right|_{\rm ss} = N \Phi(\psi^{\rm ss}) (p-q)\log(\frac{p}{q})
\end{eqnarray}
where the function $\Phi(\psi^{\rm ss})$ is some decreasing function,
i.e., $\Phi'(\psi^{\rm ss}) < 0$.
To have higher internal EP rates, the system should be more uniform
(lower $\psi^{\rm ss}$), or with a higher direct jumping rate (higher $p$). Fig.\ref{diS2_vs_psi2.eps} agrees with the conjecture that more uniform states have higher EP.

\section{VII. Summary and outlook}
In this paper, we extended the  Ehrenfest urn model with interactions and directional jumps,  allowing for detailed investigations of the non-equilibrium steady states and associated thermodynamics.
We showed that the model provides different kinds of equilibrium and non-equilibrium states.
Albeit simple the model may serve a convenient paradigm system to investigate a variety of statistical physics phenomena, ranging from equilibrium to NESS and even far from equilibrium situations.

In some situations, Landau type free energy can be constructed for NESS  near a continuous phase transition \cite{haken,Scipost}  or near the saddle-point(s) of NESS states \cite{aopingPNAS}. However, it is still highly non-trivial to construct or establish the existence of NESS free energy in general, especially in our case of coexisting NESS related by first order transitions.
On the other hand, because of the existence of
probability density $\rho^{\rm ss}(\vec x)$ at steady states, one may
define the corresponding effective potential function
$\Phi(\vec x) = - \lim_{N\rightarrow \infty} \frac{1}{N}\log\rho^{\rm ss}(\vec x)
$.
This NESS variable may reveals some NESS physical properties.
Its numerical solution will be reported elsewhere.

At high direct jumping rate and moderate coupling constant, the system is far from equilibrium and cannot
attain the steady state but limit cycle emerges instead. If the number of urns is more
than three, chaotic behavior may be possible. Such models open the  possibilities of investigating  systems with different degree of non-equilibrium  systematically.
Detailed investigations will be reported elsewhere.

\section{Acknowledgements}
We express our gratitude to the anonymous referees for their suggestions on
the improving the readability of the manuscript.
The work was supported by the Ministry of Science and Technology
of the Republic of China  under grant no. 107-2112-M-008-013-MY3 and NCTS of Taiwan.

\appendix*
\section{Appendix A: Steady State Solution of Multivariate Linear
  Fokker-Planck Equation}

The multivariate linear Fokker-Planck equation for steady state reads
\begin{eqnarray} \label{fpe1}
  -\sum_i \frac{\partial}{\partial x_i} [
    \sum_j a_{ij} x_j \rho^{\rm ss}(\vec x)
  ]
  + \frac{1}{2 N} \sum_{i j} \frac{\partial^2}{\partial x_i \partial x_j}
  [ b_{ij} \rho^{\rm ss}(\vec x)  ]  = 0 \nonumber \\
\end{eqnarray}
where $\bf a$ and $\bf b$ are constant matrices of dimension $D \times D$.
$b$ is symmetric. The natural boundary condition,
$\rho^{\rm ss}(\vec x)|_{|\vec x|\rightarrow\infty}=0$ and
$\partial_i\rho^{\rm ss}(\vec x)|_{|\vec x|\rightarrow\infty}=0$, is imposed.
The steady state was already known \cite{wang}.
In the following, we briefly outline the solution.

The form of the solution is Gaussian,
\begin{eqnarray} \label{sol01}
  \rho^{\rm ss}(\vec x) = \left(\frac{N}{\pi}\right)^{\frac{D}{2}}
  \det( -{\bf c} )^{\frac{1}{2}}
  \exp [ N \sum_{i j} c_{ij} x_i x_j  ]
\end{eqnarray}
where $\bf c$ is a symmetric matrix determined by $\bf a$ and $\bf b$.
Substitute this form into Eq.(\ref{fpe1}), we get two constraints,
\begin{eqnarray}
  {\rm tr} ({\bf a}\emph{}) &=& {\rm tr} ({\bf b c}) \label{tr1} \\
  x^{\rm t} ({\bf c a}) x &=& x^{\rm t} ({\bf c b c}) x  \label{tr2}
\end{eqnarray}
for any vector $x$.
Eq.(\ref{tr1}) is redundant (See below).

Notice that $\bf c b c$ is symmetric but $\bf c a$ is not necessary to be.
$x^{\rm t} ({\bf c a}) x = x^{\rm t} ({\bf a}^{\rm t} {\bf c}) x$ since they are both
numbers. Hence Eq.(\ref{tr2}) could be rewritten as
\begin{eqnarray}
x^{\rm t} ( {\bf c a} + {\bf a}^{\rm t} {\bf c} ) x = 2 x^{\rm t} ({\bf c b c}) x  \label{tr3}
\end{eqnarray}
for any $x$, which gives
\begin{eqnarray} \label{tr4}
  {\bf c a} + {\bf a}^{\rm t} {\bf c} = 2 {\bf c b c}
\end{eqnarray}
Take the transpose after multiplying ${\bf c}^{-1}$ from the left, one can
deduce Eq.(\ref{tr1}). Transform Eq.(\ref{tr4}) into
\begin{eqnarray} \label{tr5}
  {\bf a} {\bf c}^{-1} + {\bf c}^{-1} {\bf a}^{\rm t} = 2 {\bf b}
\end{eqnarray}
which uniquely determine $\bf c$ by noticing that the total number of independent
matrix elements of $\bf c$ is equal to the total number of independent linear equations
(both are $D(D+1)/2$).

The stability condition for
the solution in Eq.(\ref{sol01}) is negative definiteness (or equivalently,
the normalizability in infinite space). It is also equivalent to the fact that
the odd (even) order principal minor of matrix $\bf c$ is negative (positive).

\section{Appendix B: Detailed Balance in Multivariate Linear Fokker-Planck Equation}
If the steady state $\rho^{\rm ss}(\vec x)$ from the multivariate linear Fokker-Planck equation
in Eq.(\ref{fpe1}) satisfies the principle of detailed balance, {i.e.},
\begin{eqnarray} \label{db01}
  \rho^{\rm ss}(\vec x) W(\vec x';t+dt|\vec x;t)
  = \rho^{\rm ss}(\vec x') W(\vec x;t+dt|\vec x;t) \nonumber \\
\end{eqnarray}
then the steady state is also called the equilibrium state.
$W(\vec x;t+dt|\vec x';t)$ is the transition rate
from one state $\vec x'$ at time $t$
to another state $\vec x$ at time $t+dt$ \cite{risken}, which can be expressed as
\begin{widetext}
\begin{eqnarray} \label{db03}
  && W(\vec x,t+dt|\vec x',t) \nonumber \\
  &=& \left\{ 1 - (dt) \sum_{ij} a_{ij} x_j' \frac{\partial}{\partial x_i}
    + (dt) \frac{1}{2 N} \sum_{ij} b_{ij} \frac{\partial^2}{\partial x_i \partial x_j}
    + O((dt)^2)
    \right\} \delta^{(D)}(\vec x - \vec x') \nonumber \\
  &=& \exp\left[
    - (dt) \sum_{ij} a_{ij} x_j' \frac{\partial}{\partial x_i}
    + (dt) \frac{1}{2 N} \sum_{ij} b_{ij} \frac{\partial^2}{\partial x_i \partial x_j}
    \right]
  \int\frac{d^D u}{(2\pi)^D} {\rm e}^{i\sum_i u_i (x_i-x_i')}
  + O((dt)^2) \nonumber \\
  &=& \left(\frac{N}{2\pi (dt)}\right)^\frac{D}{2} ({\rm det}( {\bf b}))^{-\frac{1}{2}}
  \exp\left[
    -\frac{N}{2(dt)}\sum_{ij}({\bf b}^{-1})_{ij}
    \left(x_i-x_i'-(dt)\sum_k a_{ik} x_k'\right) \left(x_j-x_j'-(dt)\sum_k a_{jk} x_k'\right)
    \right]    \nonumber \\
\end{eqnarray}
\end{widetext}
and hence
\begin{widetext}
\begin{eqnarray} \label{db04}
\ln\left(\frac{W(\vec x;t+dt|\vec x';t)}{W(\vec x';t+dt|\vec x;t)}\right) =
\frac{N}{2}\sum_{ij}({\bf b}^{-1})_{ij} \sum_k [
    a_{jk}(x_i-x_i')+a_{ik}(x_j-x_j')
](x_k+x_k') + O(dt)
\end{eqnarray}
\end{widetext}

Notice that Eq.(\ref{db01}) is also equivalent to
\begin{eqnarray} \label{db05}
  \ln\left(\frac{\rho^{\rm ss}(\vec x)}{\rho^{\rm ss}(\vec x')}\right) =
  \ln\left(\frac{W(\vec x;t+dt|\vec x';t)}{W(\vec x';t+dt|\vec x;t)}\right)
\end{eqnarray}
by taking logarithm.
Subsitute Eq.(\ref{sol01}) and Eq.(\ref{db04}) into Eq.(\ref{db05}),
and then compare the coefficients of $x_i x_j$ and that of $x_i x_j'$ at both sides,
we have
\begin{eqnarray}
  ({\bf b}^{-1} {\bf a})_{ij} + ({\bf b}^{-1} {\bf a})_{ji} &=& 2 c_{ij} \\
  ({\bf b}^{-1} {\bf a})_{ij} - ({\bf b}^{-1} {\bf a})_{ji} &=& 0
\end{eqnarray}
in which it's matrix form is
\begin{eqnarray} \label{db06}
  {\bf c} = {\bf b}^{-1} {\bf a}
\end{eqnarray}
Here we derive the linear Fokker-Planck version of
detailed balance condition.

It is important to notice that, if ${\bf c} = {\bf b}^{-1} {\bf a}$,
then ${\bf a b} = {\bf b} {\bf a}^{\rm t}$ by direct substitution of
Eq.(\ref{db06}) into Eq.(\ref{tr5}). If we suppose
${\bf a b} = {\bf b} {\bf a}^{\rm t}$, then ${\bf c} = {\bf b}^{-1} {\bf a}$
is the solution of Eq.(\ref{tr5}). Since the solution
of Eq.(\ref{tr5}) is unique (See Appendix A), we could
draw the conclusion that Eq.(\ref{db06}) holds. ${\bf a b} = {\bf b} {\bf a}^{\rm t}$
is equivalent to ${\bf c} = {\bf b}^{-1} {\bf a}$.

Hence, ${\bf a b} = {\bf b} {\bf a}^{\rm t}$ is another equivalent statement of
detailed balance of linear Fokker-Planck equation.

\section{Appendix C: Derivation for the phase boundaries in the phase diagram}
By analysing the stability of the saddle-points as a function of $p$ and $g$, one can obtain the phase diagram for various stable states of the 3-urns model, with the phase boundary determined analytically.
First by direct calculation of the matrix $\bf a$ at the uniform saddle point of $({1\over 3},{1\over 3})$, one gets
\begin{eqnarray}
{\rm tr}({\bf a})&=&-\frac{g+3}{2}  \\
{\rm det}({\bf a})&=&{1\over {16}}[(g+3)^2+3(p-q)^2]
\end{eqnarray}
Hence the uniform NESS state is stable for $g>-3$ (${\rm tr}({\bf a})<0$ and ${\rm det} ({\bf a})>0$). In addition, the eigenvalues of $\bf a$ at the uniform NESS state can be easily calculated to give
\begin{equation}
\lambda=-\frac{g+3}{4}\pm i \frac{\sqrt{3}}{4}|p-q| \label{lamuNESS}
\end{equation}
and hence the relaxation to the uniform NESS state always has an oscillatory component.

The behavior of the equilibrium case of $p={1\over 2}$ is given in details in Ref.\cite{cheng20}.
For the non-equilibrium case of $p\neq {1\over 2}$, uniform, non-uniform NESS
and their bistable coexisting states also occur.
The phase boundary $p_{\rm c}(g)$ is determined in a similar way
by the condition of saddle-node bifurcation of a pair of stable and unstable saddle-points.
For a given $g$, $p_{\rm c}(g)$ is given by the solution of the following three equations
\begin{eqnarray}
&& A_1(x_1,x_2)=0 \\
&& A_2(x_1,x_2)=0 \\
&& \left.\frac{dx_2}{dx_1}\right|_{A_1=0}=\left.\frac{dx_2}{dx_1}\right|_{A_2=0} \label{dA}
\end{eqnarray}
for the three unknowns $p_{\rm c}$, $x_1$ and $x_2$.

\begin{figure}[htbp]
\centering
{\includegraphics[width=3in]{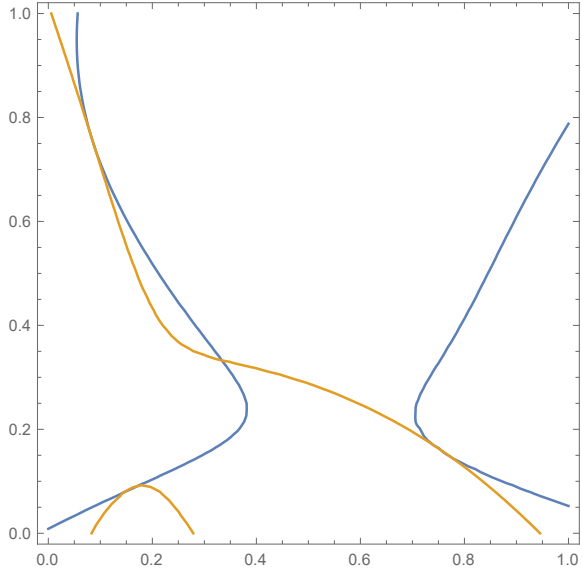}}
\caption{The parametric curves of $ A_1(x_1,x_2)=0$ and
$A_2(x_1,x_2)=0$ are shown for $p=0.8$ and $g$ near the saddle-node bifurcation. } \label{snbif}
\end{figure}

The condition of saddle-node bifurcation in Eq.(\ref{dA})
can be written out explicitly as
\begin{widetext}
\begin{eqnarray}
&& \frac{\frac{2 g [p (2 x_1+x_2-1)-x_1-x_2+1]}{\left(e^{g
   (2 x_1+x_2-1)}+1\right)^2}-\frac{e^{g (x_1+x_2)} [-g p
   x_1+g p x_2+g x_1+p+1]+p e^{2 g x_2}+e^{2 g
   x_1}}{\left(e^{g x_1}+e^{g x_2}\right)^2}+\frac{2 g [-p (2
   x_1+x_2-1)+x_1+x_2-1]+2 p-1}{e^{g (2
   x_1+x_2-1)}+1}}{\frac{g [-p (2
   x_1+x_2-1)+x_1+x_2-1]}{\left(e^{g (2
   x_1+x_2-1)}+1\right)^2}-\frac{e^{g (x_1+x_2)} [-g p
   x_1+g p x_2+g x_1+p]+p e^{2 g x_2}}{\left(e^{g
   x_1}+e^{g x_2}\right)^2}+\frac{g [p (2
   x_1+x_2-1)-x_1-x_2+1]-p+1}{e^{g (2
   x_1+x_2-1)}+1}}  \nonumber \\
&=&\frac{\frac{(p-1) e^{2 g x_1}}{\left(e^{g x_1}+e^{g
   x_2}\right)^2}+\frac{e^{g (x_1+x_2)} [(p-1) (g
   x_1+1)-g p x_2]}{\left(e^{g x_1}+e^{g
   x_2}\right)^2}+\frac{e^{g (x_1+2 x_2-1)} [p (1-g
   (x_1+2 x_2-1))+g x_2]+p}{\left(e^{g (x_1+2
   x_2-1)}+1\right)^2}}{\frac{2 g [x_2-p (x_1+2
   x_2-1)]}{\left(e^{g (x_1+2 x_2-1)}+1\right)^2}+\frac{2 p [g
   (x_1+2 x_2-1)-1
   ]-2 g x_2+1}{e^{g (x_1+2
   x_2-1)}+1}+\frac{e^{g (x_1+x_2)} \{p \sinh [g
   (x_1-x_2)]+(p-2) \cosh [g (x_1-x_2)]+g p x_1-g p
   x_2-g x_1+p-2\}}{\left(e^{g x_1}+e^{g x_2}\right)^2}} \nonumber \\
\end{eqnarray}
\end{widetext}
The phase boundary of $p_{\rm c}(g)$ for the saddle-node bifurcation together with the $g=-3$ line for stable uniform NESS are shown in the phase diagram (Fig.3 in main text), classifying the dynamics of the 3-urns model into four regimes. In the region of $g<-3$ and $p>p_{\rm c}(g)$, \emph{}there is no stable NESS state with non-steady dynamics and the system is far from equilibrium.

\begin{figure}[tbh]
 \begin{center}
   \includegraphics[width=3in]{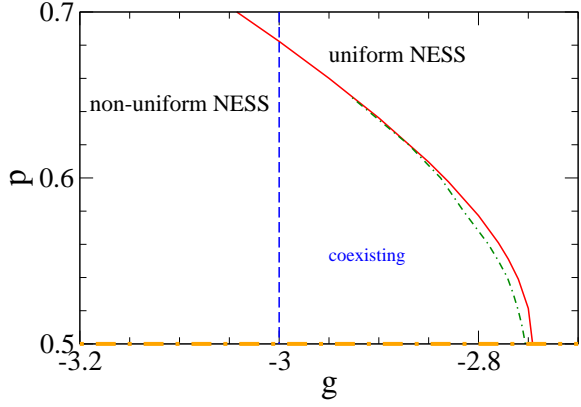}
  \end{center}
  \caption{Close-up view of the phase diagram in Fig.\ref{phase1.eps}
  of the interacting  model of three urn model near the coexistence regime.
  The dashed-dotted curve denotes the first order transition line. } \label{phasezoom.eps}
  \end{figure}

  Furthermore, the first order transition line in the coexisting regime can be analytically determined as follows. The steady state distribution near the NESS can be approximated by the Gaussian form
  $\exp(N \delta \vec x^{\rm t}{\bf c} \delta\vec x)$, where $\delta \vec x\equiv \vec x -\vec x^{\rm sp}$. In the coexisting regime, denote the relative weights of the uniform and non-uniform NESSs by $f(g)$ and $1-f(g)$, respectively, where the dependence on $g$ is written out explicitly.
  Then the steady state distribution can be expressed as
  \begin{eqnarray}
  \rho^{\rm ss}(\vec x)&=& {\cal N}^{-1}  \left( f(g) e^{N \delta {\vec x_{\rm u}}^{\rm t}
  {\bf c}_{\rm u} \delta\vec x_{u}} +[1-f(g)] e^{N \delta {\vec x_{\rm nu}}^{\rm t} {\bf c}_{\rm nu}
  \delta\vec x_{\rm nu}} \right)  \nonumber \\
  \end{eqnarray}
  where
  \begin{eqnarray}
  {\cal N} = \int d^2 x  \left(f(g) e^{N \delta {\vec x_{\rm u}}^{\rm t} {\bf c}_{\rm u} \delta\vec x_{\rm u}} +[1-f(g)] e^{N \delta {\vec x_{\rm nu}}^{\rm t} {\bf c}\emph{}_{\rm nu} \delta\vec x_{\rm nu}} \right)
  \nonumber \\
  \end{eqnarray}
  is the normalization factor, and the subscripts $\rm u$ and $\rm nu$ denote uniform and non-uniform NESS, respectively. The ensemble average of the steady state flux can be computed using saddle point approximation to give
  \begin{eqnarray}
  \langle K^{\rm ss}(g) \rangle \simeq \chi(g)  K^{\rm ss}_{\rm u}
  + [1-\chi(g)] K^{\rm ss}_{\rm nu}(g)  \label{meanKss}
  \end{eqnarray}
  where
  \begin{eqnarray}
   \chi(g) = \frac{f(g)}{f(g)+[1-f(g)]\sqrt{\frac{\det ({\bf c}_{\rm u}(g)) }{\det ({\bf c}_{\rm nu}(g))}}}
  \end{eqnarray}
  At the first order transition point $g_{\rm t}$, $f(g_{\rm t})={1\over 2}$ and the R.H.S. of Eq.(\ref{meanKss}) reduces to
  \begin{eqnarray}
  \phi(g) K^{\rm ss}_{\rm u} +[1-\phi(g)] K^{\rm ss}_{\rm nu}(g)  \label{Kss1st}
  \end{eqnarray}
  where
  \begin{eqnarray}
  \phi(g) = \frac{1}{1+\sqrt{\frac{\det ({\bf c}_{\rm u}(g)) }{\det ({\bf c}_{\rm nu}(g))}}}
  \end{eqnarray}
 which can be analytically calculated as a function of $g$. Thus by numerically computing
 $\langle K^{\rm ss}(g) \rangle$ using the numerical solution of Eq.(2),
 $g_{\rm t}$ can be obtained from the intersection of the curves of $\langle K^{\rm ss}(g) \rangle$ and  Eq.(\ref{Kss1st}). For given values of $p$ in the coexistence regime, $g_{\rm t}$ is obtained theoretically from the above manner and the result of the first order transition line is shown in
 Fig.\ref{phasezoom.eps}. The first order line is rather close to the stability boundary of the non-uniform NESS indicates that the non-uniform NESS dominates over the uniform NESS in the coexistence regime. This echoes with the observation in Fig.\ref{lambdanuNESS.eps} that the eigenvalues of $\bf a$ of the non-uniform NESS are much more negative than that of (the real part) the uniform NESS unless $p$ is very close to the stability boundary, indicating that the non-uniform NESS is a strong attractor than that of the uniform NESS in most of the coexistence regime.



\begin{thebibliography}{}
\bibitem{saddle}
In equilibrium statistical mechanics, when $N$ is large, in computing the partition function or any average physical
variables, the method of steepest decent (one of the popular asymptotic methods) is applied.
In this way, the physical varaibles we consider will be extended from real to complex plane.
The "optimal" point in this complex plane is not locally maximum or minimum, but a saddle point
(Chapter 9 in Ref.[2]). This saddle point in the complex plane corresponds to the fixed point
in the parameter space of the Fokker-Planck Equation in Eq.(\ref{fkeq02}).
\bibitem{wkb}
The WKB method is adopted in Section 6.6.7 from Ref.\cite{risken}, which is the usual analytical method to solve the
Fokker-Planck Equation with small diffusion coefficient (large $N$ limit in our case).
\end{thebibliography}
\end{document}